\documentclass{article}

\usepackage{spconf,amsmath,graphicx, booktabs}
\usepackage{amssymb, multirow, color,cite}
 \usepackage[colorlinks,
             linkcolor=red,
             anchorcolor=blue,
            citecolor=green]{hyperref}
 \DeclareMathOperator*{\argmin}{argmin}

\title{EXPLORING STRUCTURAL SPARSITY IN NEURAL IMAGE COMPRESSION}
\name{Shanzhi Yin\textsuperscript{1}, Chao Li\textsuperscript{1},   Wen Tan\textsuperscript{1}, Youneng Bao\textsuperscript{1}, Yongsheng Liang\textsuperscript{1}, Wei Liu\textsuperscript{2}}
\address{\textsuperscript{1}Harbin Institute of Technology, Shenzhen   \qquad  \textsuperscript{2}PengCheng Laboratory}
\begin{document}
%
\maketitle
\begin{abstract}
Neural image compression have reached or out-performed traditional methods (such as JPEG, BPG, WebP). However, their sophisticated network structures with cascaded convolution layers bring heavy computational burden for practical deployment.  In this paper, we explore the structural sparsity in neural image compression network to obtain real-time acceleration without any specialized hardware design or algorithm. We propose a simple plug-in adaptive binary channel masking(ABCM) to judge the importance of each convolution channel and introduce sparsity during training. During inference, the unimportant channels are pruned to obtain slimmer network and less computation.  We implement our method into three neural image compression networks with different entropy models to verify its effectiveness and generalization, the experiment results show that up to 7$\times$ computation reduction and 3$\times$ acceleration can be achieved with negligible performance drop. 

\end{abstract}

\section{Introduction}
\label{sec:intro}

Image compression is an essential and fundamental research topic in signal processing and computer vision. It minimizes the required bits for image coding and reconstruction errors for image decoding at the same time. Recently, neural image compression networks have shown great potential to boost the Rate-Distortion optimization and achieved state-of-the-art performance compared to traditional methods
\cite{minnen2018joint,cheng2020learned,chen2021end,cui2021asymmetric}. However, these advanced networks are commonly based on complex convolution calculations. Their model complexity can reach more than 50G FLOPs\cite{cheng2019deep}, which brings heavy computational overhead for practical hardware deployment and is unfriendly to circumstances like portable electrical devices or edge computing.
\begin{figure}[t]

    \centering
    \includegraphics[height=2.6cm, width=8.5cm]{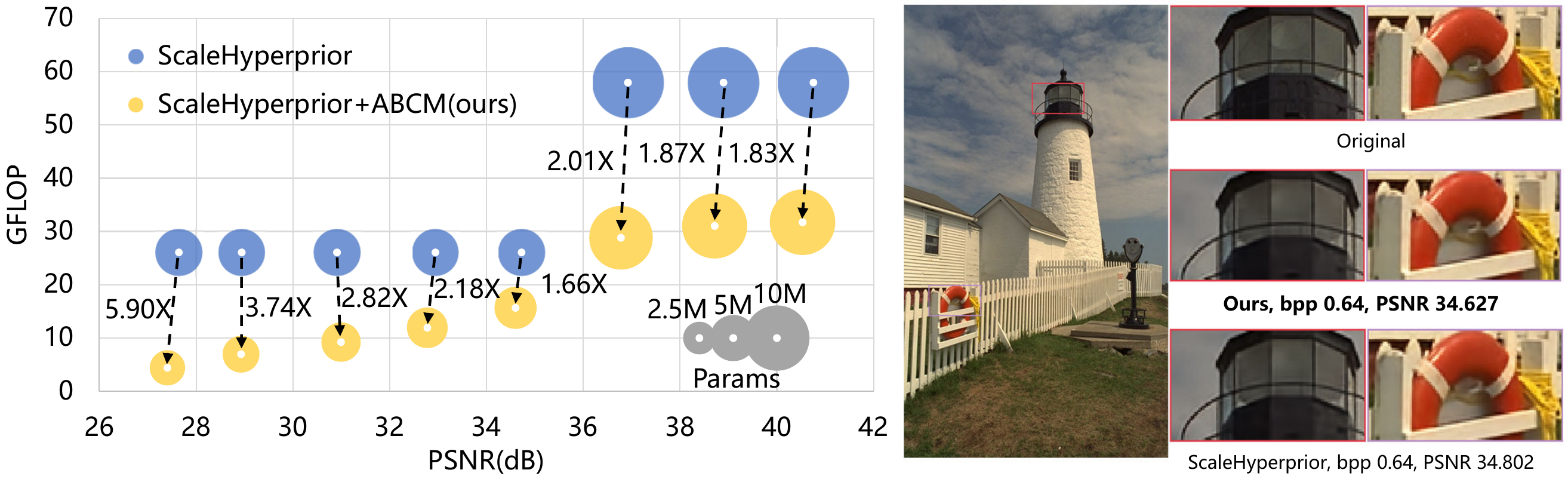}
    \caption{Calculation (up to 5.9$\times$) and parameters (up to 1.7$\times$)  saving on ABCMs embedded ScaleHyperprior models of 8 different quality while performance drop is negligible(left); Our method shows no obvious visual qualitative degradation on the reconstructed image(right).    }
    
    \label{fig:1}

\end{figure}

To simplify the network structure or reduce the computation of convolution neural network, many researchers have developed network sparsification or dynamic inference techniques.  BN parameters\cite{liu2017learning} and  global probabilistic constraint\cite{zhou2021effective} are utilized to prune ineffective structures. Stochastic sampling\cite{xie2020spatially} and channel-dimension correlation\cite{hua2019channel}  are used to judge the important area of current feature maps and implement spatially dynamic  inference.  However, these methods either lack implementation on image compression task or need specialized hardware programming to achieve real-time acceleration. 
 
 In the meantime, sparsity in neural image compression remains an open and less researched problem. Some researchers only focus on the simplification of the neural image decoder with regulation technique\cite{johnston2019computationally}, multi-branch module\cite{guo2021cbanet}, or introduce complicated training algorithm\cite{yang2021slimmable} with slimmable network\cite{yu2018slimmable}. In our previous work\cite{yin2021universal}, we obtain spatial-dimension sparse feature map of NIC models in a simple but effective way, but it needs dynamic convolution and is accordingly unfriendly to real-time acceleration. On the other hand, channel-dimension sparsity can be leveraged to obtain slimmer network structure and maintain static inference, which would be practical for direct hardware acceleration.

\begin{figure}[t]

    \centering
    \includegraphics[width=8.5cm]{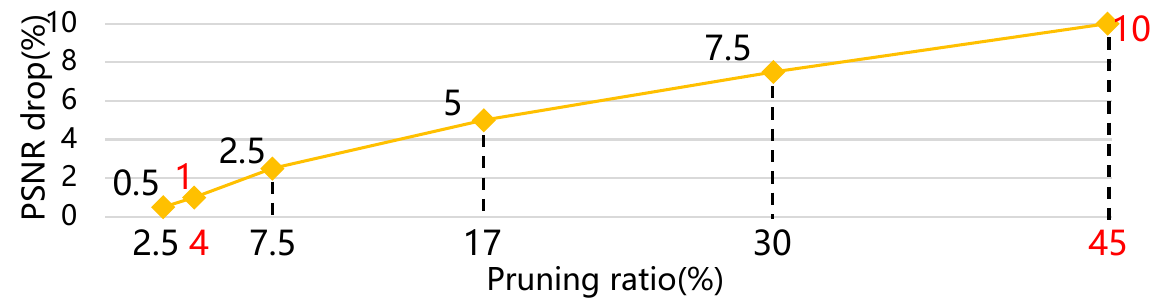}
    \caption{PSNR drop vs. pruning ratio on trained model}
    \label{fig:23}

\end{figure}

 To get a slimmer network structure explicitly, we first investigated the channel redundancy in a trained neural image compression(NIC) model. As discussed in \cite{ijcai2020-65}, the importance of different channels in NIC models varies, removing each of them can result in different reconstruction results.   In our experiments, we search layer by layer in main encoder of a trained ScaleHyperprior model\cite{balle2018variational}. In each layer, we iteratively search and prune one channel that causes the minimum PSNR drop of reconstruction until no channel can be pruned without PSNR drop lager than a certain threshold. Then, we move to the next layer. The layers are searched from  input end to the latent end and from decoder-side to encoder-side on each end, corresponding to their potential influence to the reconstruction quality. We set the maximum PSNR drop threshold to different values and see the corresponding pruning ratio that the trained model can achieve. The results in Figure.\ref{fig:23} shows that a relatively small  PSNR drop value like 1\% can only allow a small pruning ratio of no more than 5\%, while a pruning ratio of nearly 50\% can lead to unacceptable performance drop of more then 10\%.
 
Figure.\ref{fig:23} indicates that structural channel sparsity is insufficient in a trained NIC model and needs to be introduced during training procedure. This inspires us to propose adaptive binary channel masking(ABCM) as shown in Fig.\ref{fig:2}. We use a learnable vector to automatically judge the importance of every convolution output channels. By gating the learnable vector, we can obtain a binary channel-dimension mask, which is then masked on the each output channel to set the unimportant channels to zero. By doing so, we introduce structural sparsity during the training procedure and force the network to gather useful information to important channels decided by ABCM as much as possible. During inference, the zeroed channels are pruned to get slimmer network structure and real-time acceleration subsequently. 
 
 We implement our method on three NIC models in a plug-in manner and conduct comprehensive experiments to prove the effectiveness of our method. The effectiveness on ScaleHyperprior model shown in Fig.\ref{fig:1}. There is no obvious performance degradation on the reconstructed image, while the parameter and computation saving is quite impressive.

As far as we know, this may be the first work exploring the structural sparsity in neural image compression for real-time interference acceleration. Our contribution are summarized as followed:

$\bullet$ We propose a simple but effective ABCM method that can judge the importance of convolution channels and  introduce sparsity during training procedure. 

$\bullet$ We implement our ABCM in a plug-in manner and explore the structural sparsity in NIC models to achieve real-time inference acceleration.

$\bullet$ Comprehensive experiment results show that up to 7$\times$ computation reduction and 3$\times$ acceleration can be achieved with negligible performance drop. 


\section{Proposed Method}
\label{sec:method}

\subsection{Adaptive binary channel masking}
 In\cite{yin2021universal}, we obtain sparse feature map from NIC convolution output by spatial-wisely judging the significance of each element and zero unimportant locations. In this work ,we focus on the importance of different convolution channels, which is adaptively learned by ABCMs during the training procedure.
 
\begin{figure}[t]

    \centering
    \includegraphics[width=8.5cm]{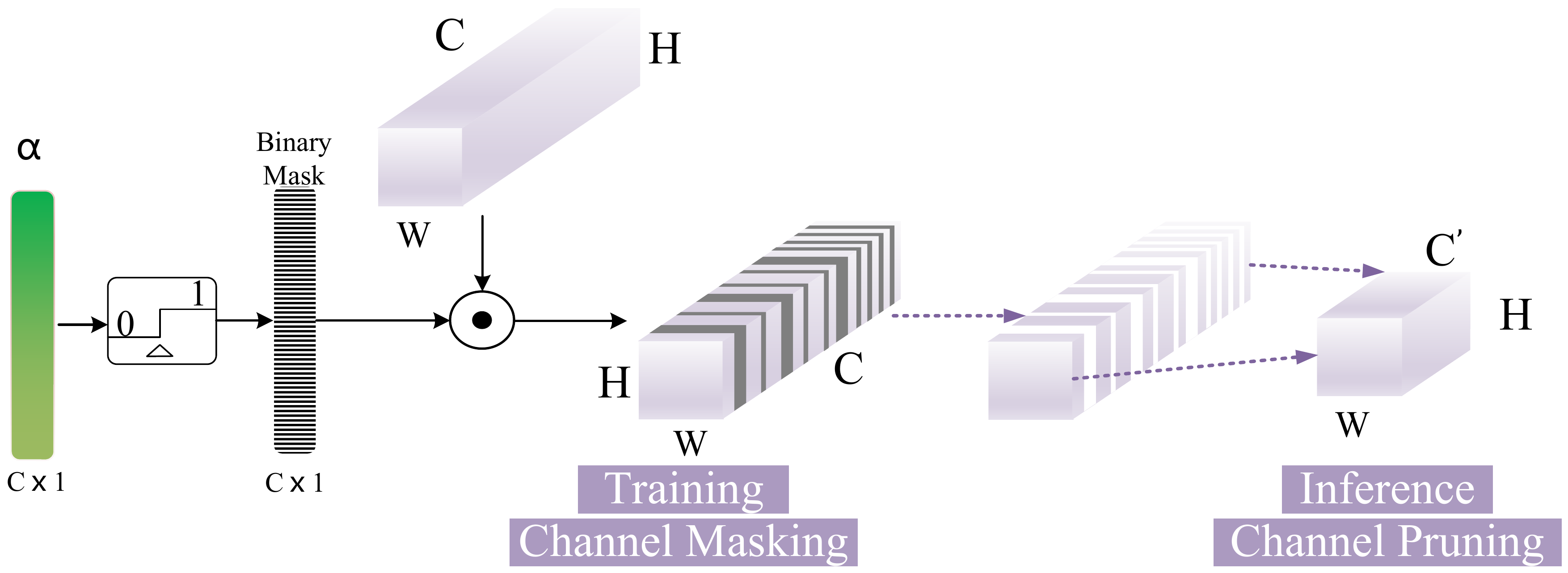}
    \caption{Proposed adaptive binary channel masking}
    \label{fig:2}

\end{figure}

 As shown in Fig.\ref{fig:2}, we denote a learnable important vector as $\alpha$, which is supposed to automatically judge the influence of each channel to the model performance during the training procedure. Then, the importance vector is gated by a simple gating function to obtain a binary channel mask, the gating function can be denoted as:
\begin{equation}\label{gt}
gt(\alpha) =
\begin{cases}
1 & \text{if }\alpha \ge 0,\\
0 & \text{if } otherwise,\\
\end{cases} 
\end{equation}For this gating function is not differentiable, Sigmoid function $\sigma(\alpha)=\frac{1}{1+e^{-\epsilon \alpha}}$ is used to replace it during the training process\cite{hua2019channel}.
If we denote the output feature map of convolution layer as $x$, ABCM module can set the unimportant channels to zero and its output can be denoted as:
\begin{equation}\label{xmaks}
x_{mask} = x\odot gt(\alpha)
\end{equation}
in which $\odot$ denotes the channel-wise product and $x_{mask}$  is the channel-dimension sparse feature map. During inference, these zeroed channels in every layer are pruned to get a slimmer network structure with parameters only from effective channels. As the computation of convolution operation is related to the number of its input and output channels, faster inference can be realized with slimmer network.

It should be noted that, during inference, ABCM modules are only used to guide the network pruning and are not used in the consequent slimmer network. The function of ABCM is two-fold: firstly, it introduces sparsity  by setting certain channels to zero and forces the network to minimize the effectiveness of these channels during training, so that the original performance of the models can be reserved as much as possible; secondly, by using binary channel mask to introduce sparsity during training, the performance of slimmer model can maintain the same before and after the channel pruning.
\subsection{Implementation and optimization}
\begin{figure}[t]

    \centering
    \includegraphics[width=8.5cm, height=4.5cm]{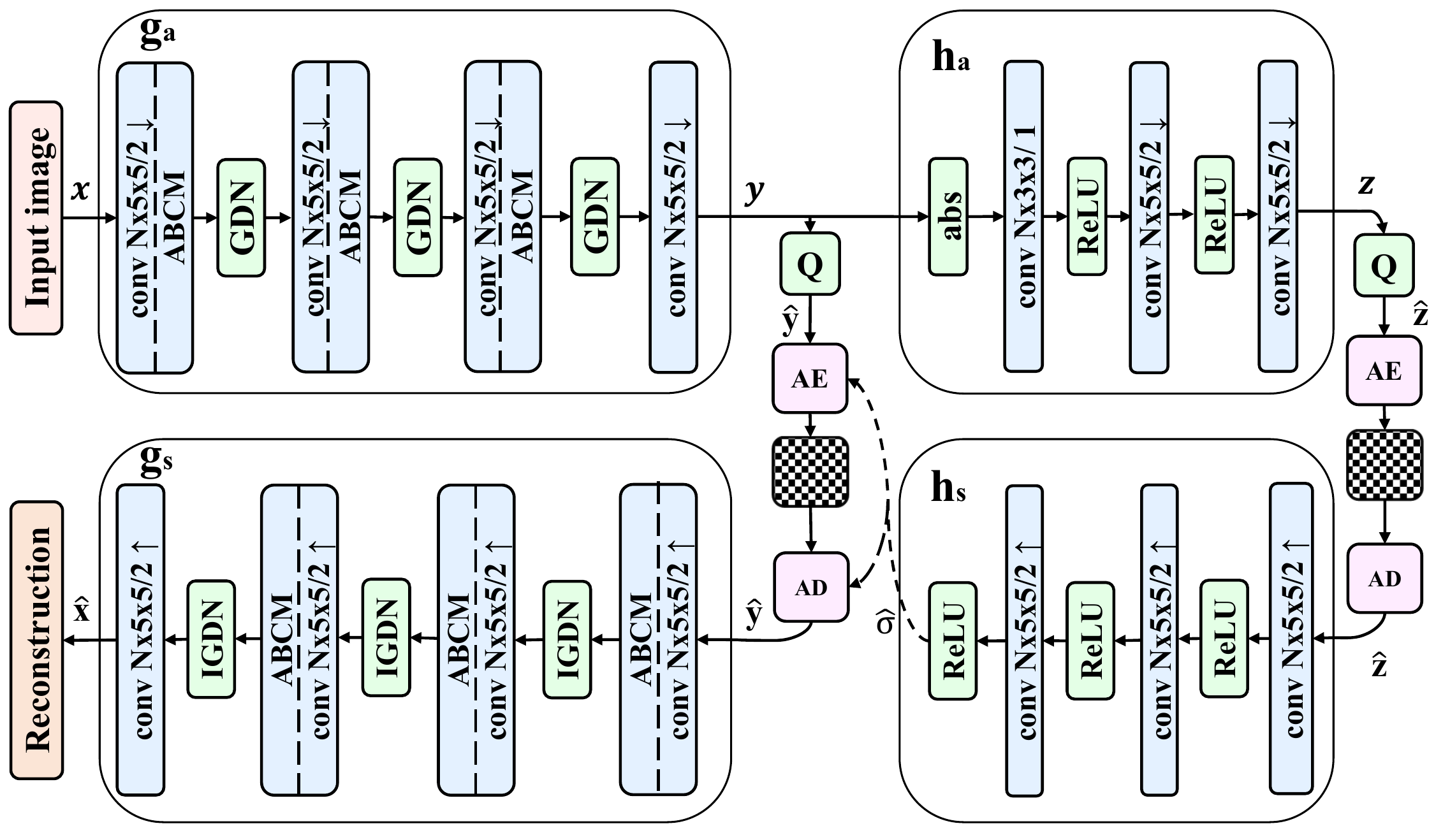}
    \caption{Implementation on ScaleHyperprior model}
    \label{fig:3}

\end{figure}
To illustrate the effectiveness of our method, we implement ABCM into three existing NIC models\cite{Balle2016,balle2018variational,minnen2018joint} with different entropy models, denoted as Factorizedprior model, ScaleHyperprior model, JointAutoregressive model respectively. 
In this section, we use ScaleHyperprior model as an example to explain the implement details and optimization formulation of our method.

 As shown in Fig.\ref{fig:3}, we plug totally 6 ABCM modules after convolution layers. The last layer of the main encoder and main decoder are unchanged to maintain the channel number of the network input and entropy coding. Besides, the hyper codec remains unchanged, because we notice that the hyper parameters are significant for keeping an accurate entropy model. If we introduce sparsity in hyper codec, the computational saving cannot compensate for the corresponding performance degradation.
 
 After adopting our method into the original NIC models, it needs to be trained in a end-to-end manner. Similar to most image compression problem, two loss terms rate $R$ and distortion $D$ need to be optimized. As the actual arithmetic coding is bypassed\cite{Balle2016}, the rate term is given by the cross entropy of the estimated distribution of y and the its actual distribution:
 \begin{equation}\label{rate}
R(\hat{y};\theta,\phi) = \mathbb{E}_{\hat{y}\sim p_{y}}\{log_2q_{y}[Q(y)]\}
\end{equation}
in which, $\theta$, $\phi$ denote the parameters of codecs and ABCM modules respectively, Q represents the quantization process, $p$ and $q$ are actual and predicted distribution of image data respectively. In our work, the distortion between the input image and reconstructed image is the mean square error(mse) measured on the test set:
\begin{equation}\label{distortion}
D(x,\hat{x};\theta,\phi) = \mathbb{E}_{x\sim p_{x}}[||x-\hat{x}||^2]
\end{equation}
Besides, we set a additional sparsity regulation term based on our  ABCM modules to ensure the channel-dimension sparsity during the training and achieve the comprehensive trade-off between rate, distortion and sparsity. As discussed in section.\ref{sec:method}, we denote the importance vector in the $i$ th ABCM modules as $\alpha_{i}$, gating function as $gt$, the corresponding sparsity term $s_{i}$ can be expressed as:
\begin{equation}\label{sparsity}
s_{i} = \frac{||gt(\alpha_{i})||_{1}}{C_{i}}
\end{equation}
in which, $||\cdot||_{1}$ denotes the L1-norm and $C_{i}$ denotes the channel number of ABCM module. The final optimization problem with $n$ ABCM modules can be formulated as:

\begin{equation}\label{loss}
\mathop{\argmin_{\theta,\phi}}[R+\lambda D+\gamma \frac{1}{n} \sum_{i=1}^n s_{i}]
\end{equation}
in which, $R$, $D$, $s_{i}$ are defined by equation(\ref{rate})(\ref{distortion})(\ref{sparsity}) respectively, $\lambda$ and $\gamma$ are hyper parameters to control the trade-off between rate, distortion and sparsity. When tuning these two hyper parameters, a proper level of sparsity can be achieved with acceptable or even negligible performance degradation. 
\begin{table*}[t]
  \centering
  \caption{The PSNR, Parameters and FLOP reduction of models on Kodak set}
  \renewcommand\arraystretch{0.9}
    \begin{tabular}{c|c|cccccccc}
    \toprule
    \multirow{2}{*}{Model} & \multirow{2}{*}{Reduction} & \multicolumn{8}{c}{Quality} \\
\cline{3-10}          &       & 1     & 2     & 3     & 4     & 5     & 6     & 7     & 8 \\
 \bottomrule
 \multirow{3}{*}{Factorizedprior} & PSNR(\%) & 0.739     &  0.391     &   0.990    & 0.552     &   0.440    & 0.728     &    0.298  & 0.223 \\
    & Params &  \textbf{5.63$\times$}     & \textbf{ 3.80$\times$}     &\textbf{2.81$\times$ }     &   \textbf{2.13$\times$}    &     \textbf{1.70$\times$}  &\textbf{ 1.60$\times$}      &\textbf{ 1.40$\times$}      &\textbf{1.31$\times$ }\\
          & FLOP (with ABCMs)&   \textbf{7.17$\times$}    & \textbf{4.35$\times$}      &\textbf{ 3.01$\times$}      & \textbf{2.17$\times$}      &\textbf{1.61$\times$}       &\textbf{2.13$\times$}       &\textbf{1.83$\times$}       &\textbf{1.72$\times$} \\

    \hline
\multirow{4}{*}{ScaleHyperprior} & PSNR(\%) & 0.857     &  0.052     &   0    & 0.504     &   0.306    & 0.393     &    0.473  & 0.552 \\
    & Params &  \textbf{1.73$\times$}     & \textbf{ 1.60$\times$  }   &\textbf{1.39$\times$}      & \textbf{  1.30$\times$}    &    \textbf{ 1.22$\times$}  & \textbf{1.25$\times$   }   &\textbf{ 1.19$\times$}      &\textbf{1.17$\times$} \\
          & FLOP (with ECGs\cite{yin2021universal}) &   2.54$\times$   & 2.86$\times$      &2.60 $\times$      & 2.54$\times$      &2.54$\times$      &2.07$\times$       &2.14$\times$      &2.03$\times$ \\
          & FLOP (with ABCMs)&   \textbf{5.90$\times$}    & \textbf{3.74$\times$}      &\textbf{ 2.82$\times$}      & \textbf{2.18$\times$}      &\textbf{1.66$\times$}       &\textbf{2.01$\times$}       &\textbf{1.87$\times$}       &\textbf{1.83$\times$} \\

     \hline
    \multirow{4}{*}{JointAutoregressive} & PSNR(\%) & 1.17     &  0.963     &   0.553    & 0.662     &   0.402    & 0.601     &    0.667  & 0.451 \\
    & Params &  \textbf{1.49$\times$}     &  \textbf{1.43$\times$ }    &\textbf{1.35$\times$}      &   \textbf{1.30$\times$}    &    \textbf{ 1.13$\times$}  & \textbf{1.09$\times$  }    & \textbf{1.08$\times$}      &\textbf{1.05$\times$ }\\
          & FLOP (with ECGs\cite{yin2021universal}) &   2.34$\times$   & 2.50$\times$      &2.56 $\times$      & 2.68$\times$      &2.33$\times$      &2.12$\times$       &2.12$\times$      &2.24$\times$ \\
          & FLOP (with ABCMs)&   \textbf{6.39$\times$}    & \textbf{4.77$\times$}      &\textbf{ 3.76$\times$}      & \textbf{2.80$\times$}      &\textbf{2.24$\times$}       &\textbf{1.89$\times$}       &\textbf{1.78$\times$}       &\textbf{1.53$\times$} \\

    
    \bottomrule
    \end{tabular}%
  \label{tab:table1}%
\end{table*}%

\begin{table*}[t]
  \centering
  \caption{The real-time acceleration for 768$\times$512 input image on different devices }
  \renewcommand\arraystretch{0.9}
  \setlength{\tabcolsep}{2.6mm}{
    \begin{tabular}{c|c|cccccccc}
    \toprule
    \multirow{2}{*}{Model} & \multirow{2}{*}{Devices} & \multicolumn{8}{c}{Quality} \\
\cline{3-10}          &       & 1     & 2     & 3     & 4     & 5     & 6     & 7     & 8 \\
    \hline
    \multirow{2}{*}{Factorizedprior} &Intel  Xeon  Gold 5118 &  3.40$\times$     &  2.67$\times$      & 2.07$\times$       &   1.91$\times$     &     1.39$\times$   & 1.65$\times$       & 1.46$\times$      & 1.35$\times$  \\
       
      & NVIDIA Tesla V100 &  1.30$\times$     &  1.28$\times$      & 1.16$\times$       &   1.13$\times$     &     1.09$\times$   & 1.78$\times$       & 1.53$\times$       & 1.50$\times$  \\
    \hline
    \multirow{2}{*}{ScaleHyperprior} 
          & Intel Xeon Gold 5118 &  3.12$\times$     &  2.19$\times$      & 1.96$\times$       &   1.90$\times$     &     1.38$\times$   & 1.45$\times$       & 1.39$\times$       & 1.37$\times$  \\
      & NVIDIA Tesla V100 &  1.31$\times$     &  1.29$\times$      & 1.18$\times$       &   1.14$\times$     &    1.10$\times$   & 1.75$\times$       & 1.61$\times$       & 1.58$\times$  \\
       \hline
    \multirow{2}{*}{JointAutoregressive} & Intel Xeon Gold 5118 &  2.83$\times$     &  2.60$\times$      & 1.99$\times$       &   1.62$\times$     &     1.45$\times$   & 1.40$\times$       & 1.41$\times$       & 1.26$\times$  \\
          & NVIDIA Tesla V100 &  2.33$\times$     &  2.12$\times$      & 2.08$\times$       &   1.91$\times$     &     1.81$\times$   & 1.57$\times$       & 1.53$\times$       & 1.19$\times$  \\

    \bottomrule
    \end{tabular}}%
  \label{tab:table2}%
\end{table*}%
\section{EXPERIMENTS}
\label{sec:experiments}
We use CLIC dataset\footnote{The CLIC Dataset can be accessed at http://compression.cc/} to train our models for 100 epochs. We adopted CompressAI\cite{begaint2020compressai} re-implantation of NIC models and followed most of their training settings, except that we reduce learning rate by 0.5 only once after 64 epochs, i.e. approximately 1 million steps.
All trained models are evaluated on Kodak dataset\footnote{The Kodak Dataset can be accessed at http://r0k.us/graphics/kodak/}. 
\subsection{Ablation study}
\begin{figure}[t]

    \centering
    \includegraphics[width=8.5cm]{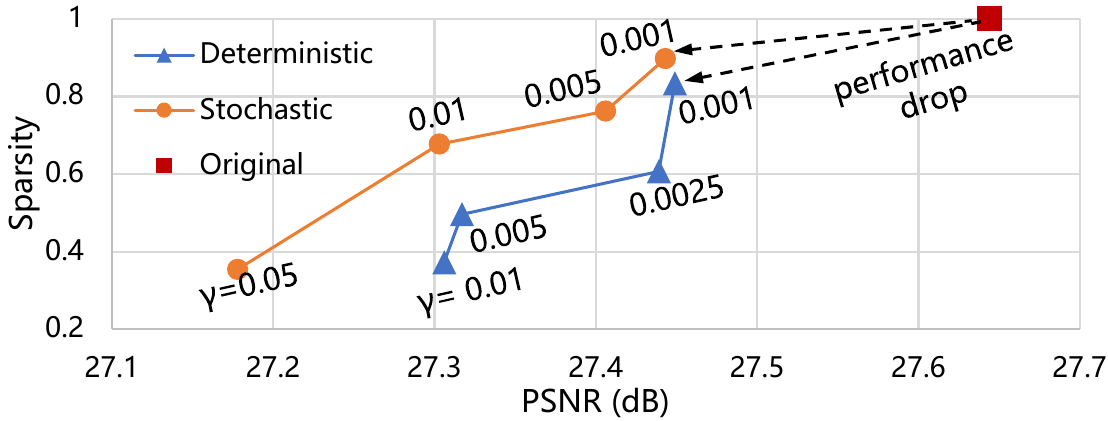}
    \caption{Ablation study on mask generation and sparsity regulation term, sparsity denotes the ratio of effective channels.}
    \label{fig:4}

\end{figure}
Before we implement our method, we did ablation study on two aspects, using ScaleHyperprior quality=1 model.

\textbf{Mask generation}. As several work has illustrated the effectiveness of treating network sparsity as a probabilistic event\cite{wang2021exploring,zhou2021effective}, the importance vector can be either a learnable parameter itself, or a Bernoulli sampling from a $2\times C$ learnable parameter with reparameterization trick\cite{jang2016categorical}. We denote these two methods as Deterministic and Stochastic.

\textbf{Sparsity regulation term}. The hyper parameter $\gamma$  controls the trade-off between network sparsity and performance. Choosing a appropriate $\gamma$  can contribute to the comprehensive optimization in equation(\ref{loss}).

As we tune the value of $\gamma$, the bpp remains 0.14, while the PSNR($10log_{10}\frac{255^2}{mse}$) and average sparsity of all ABCM modules on test set are shown in Fig.\ref{fig:4}.  We can see that the deterministic mask generation can achieve relatively higher sparsity with less performance drop. To maintain acceptable performance drop and obtain network sparsity as much as possible, we choose deterministic mask generation and set $\gamma=0.01$ in subsequent experiments.


\subsection{R-D and efficiency performance}

We implement our method in three NIC models to build efficient models with promising saving in both parameters and computation. When embedded with ABCMs, we observe that the bit-rate of our models fluctuate around their original ones. For fair comparison,we fine-tuned $\lambda$ and trained one more epoch so that our bit-rates are the same (accurate to two decimal places) as their original counterparts. 

Table.\ref{tab:table1} shows the parameters and FLOP reduction of efficient models obtained by ABCMs guided pruning. We can see that up to 5.63$\times$, 1.73$\times$, 1.49$\times$ parameters saving and  up to 7.17$\times$, 5.90$\times$, 6.39$\times$ FLOP reduction can be achieved in three models with PSNR drop around 0.5\%. It should be noted that, such efficiency is achieved under hyperparameter $\gamma=0.01$, when tuning its value, the trade-off between performance and efficiency can be arbitrary.

The results show that channel redundancy commonly exist in neural image compression network. Under the same channel number configuration, lower the quality is (smaller the hyperparameter $\lambda$ is), sparser the network can be. Even with the highest quality model, the FLOP can be reduced by nearly 2 times. Compared to ECG modules in \cite{yin2021universal}, our ABCMs perform much better in low quality model and can be directly utilized for real-time acceleration.

To measure the real-time acceleration of efficient models on both CPU and GPU, we chose Kodim.19 in Kodak set as the input image. Considering the warm-up and multi-threading of the devices, we take ten rounds of inference as warm-up and take another ten rounds as formal tests, then record the average duration of each round. 

Table.\ref{tab:table2} shows the real-time acceleration (excluding data loading/writing and arithmetic coding) on a NVIDIA Tesla V100 GPU and Intel(R) Xeon(R) Gold 5118 CPU, which corresponds to the FLOP reduction in Table.\ref{tab:table1}. We can see that  up to 3.40$\times$, 3.12$\times$, 2.83$\times$ real-time CPU acceleration and  up to 1.30$\times$, 1.31$\times$, 2.33$\times$ real-time GPU acceleration can be achieved in three models. Such acceleration would be more obvious and beneficial on big data or continuous processing.

\section{Conclusion}
\label{sec:conclusion}
In this paper, we explore the channel-dimension structural sparsity in existing neural image compression models to obtain real-time acceleration. We implementing our plug-in adaptive binary channel masking in three different NIC models to verify its effectiveness and generalization. Comprehensive experiment results show that 1.2-3.4$\times$ inference acceleration and 1.5-7.1$\times$ parameter saving can be achieved with less than 1\% performance drop. Potential future work includes better joint optimization, complexity-rate control.

\vfill\pagebreak

\bibliographystyle{IEEEbib}
\bibliography{refs}
\newpage
\section{Appendix}

\begin{table}[htbp]\tiny
  \centering
  \caption{Detailed channel configuration of efficient models}
    \begin{tabular}{|cc|cc|cc|cc|cc|cc|cc|cc|cc|cc|cc|}
    \toprule
    \multicolumn{22}{|c|}{\textbf{ScaleHyperprior}} \\
    \midrule
    \multirow{2}[2]{*}{\textbf{Module}} & \multirow{2}[2]{*}{\textbf{Layer}} & \multicolumn{2}{c|}{\textcolor[rgb]{ 1,  0,  0}{\textbf{channel (low)}}} & \multicolumn{2}{c|}{\textcolor[rgb]{ 1,  0,  0}{\textbf{channel (high)}}} & \multicolumn{2}{c|}{\textcolor[rgb]{ 1,  0,  0}{\textbf{ (Q1)}}} & \multicolumn{2}{c|}{\textcolor[rgb]{ 1,  0,  0}{\textbf{(Q2)}}} & \multicolumn{2}{c|}{\textcolor[rgb]{ 1,  0,  0}{\textbf{(Q3)}}} & \multicolumn{2}{c|}{\textcolor[rgb]{ 1,  0,  0}{\textbf{ (Q4)}}} & \multicolumn{2}{c|}{\textcolor[rgb]{ 1,  0,  0}{\textbf{(Q5)}}} & \multicolumn{2}{c|}{\textcolor[rgb]{ 1,  0,  0}{\textbf{ (Q6)}}} & \multicolumn{2}{c|}{\textcolor[rgb]{ 1,  0,  0}{\textbf{(Q7)}}} & \multicolumn{2}{c|}{\textcolor[rgb]{ 1,  0,  0}{\textbf{(Q8)}}} \\
          &       & in    & out   & in    & out   & in    & out   & in    & out   & in    & out   & in    & out   & in    & out   & in    & out   & in    & out   & in    & out  \\
    \midrule
    \multirow{7}[2]{*}{\textbf{ga}} & conv  & 3     & 128   & 3     & 192   & 3     & 30    & 3     & 38    & 3     & 50    & 3     & 66    & 3     & 95    & 3     & 111   & 3     & 128   & 3     & 133 \\
          & GDN   & 128   & 128   & 192   & 192   & 30    & 30    & 38    & 38    & 50    & 50    & 66    & 66    & 95    & 95    & 111   & 111   & 128   & 128   & 133   & 133 \\
          & conv  & 128   & 128   & 192   & 192   & 30    & 39    & 38    & 48    & 50    & 71    & 66    & 90    & 95    & 107   & 111   & 146   & 128   & 166   & 133   & 176 \\
          & GDN   & 128   & 128   & 192   & 192   & 39    & 39    & 48    & 48    & 71    & 71    & 90    & 90    & 107   & 107   & 146   & 146   & 166   & 166   & 176   & 176 \\
          & conv  & 128   & 128   & 192   & 192   & 39    & 48    & 48    & 63    & 71    & 86    & 90    & 94    & 107   & 88    & 146   & 144   & 166   & 166   & 176   & 174 \\
          & GDN   & 128   & 128   & 192   & 192   & 48    & 48    & 63    & 63    & 86    & 86    & 94    & 94    & 88    & 88    & 144   & 144   & 166   & 166   & 174   & 174 \\
          & conv  & 128   & 192   & 192   & 320   & 48    & 192   & 63    & 192   & 86    & 192   & 94    & 192   & 88    & 192   & 144   & 320   & 166   & 320   & 174   & 320 \\
    \midrule
    \multirow{7}[2]{*}{\textbf{gs}} & conv  & 192   & 128   & 320   & 192   & 192   & 81    & 192   & 88    & 192   & 117   & 192   & 117   & 192   & 126   & 320   & 191   & 320   & 189   & 320   & 178 \\
          & GDN   & 128   & 128   & 192   & 192   & 81    & 81    & 88    & 88    & 117   & 117   & 117   & 117   & 126   & 126   & 191   & 191   & 189   & 189   & 178   & 178 \\
          & conv  & 128   & 128   & 192   & 192   & 81    & 41    & 88    & 57    & 117   & 63    & 117   & 77    & 126   & 89    & 191   & 118   & 189   & 119   & 178   & 128 \\
          & GDN   & 128   & 128   & 192   & 192   & 41    & 41    & 57    & 57    & 63    & 63    & 77    & 77    & 89    & 89    & 118   & 118   & 119   & 119   & 128   & 128 \\
          & conv  & 128   & 128   & 192   & 192   & 41    & 40    & 57    & 62    & 63    & 70    & 77    & 80    & 89    & 94    & 118   & 124   & 119   & 128   & 128   & 121 \\
          & GDN   & 128   & 128   & 192   & 192   & 40    & 40    & 62    & 62    & 70    & 70    & 80    & 80    & 94    & 94    & 124   & 124   & 128   & 128   & 121   & 121 \\
          & conv  & 128   & 3     & 192   & 3     & 40    & 3     & 62    & 3     & 70    & 3     & 80    & 3     & 94    & 3     & 124   & 3     & 128   & 3     & 121   & 3 \\
    \midrule
    \multicolumn{1}{r}{} & \multicolumn{1}{r}{} &       & \multicolumn{1}{r}{} &       & \multicolumn{1}{r}{} &       & \multicolumn{1}{r}{} &       & \multicolumn{1}{r}{} &       & \multicolumn{1}{r}{} &       & \multicolumn{1}{r}{} &       & \multicolumn{1}{r}{} &       & \multicolumn{1}{r}{} &       & \multicolumn{1}{r}{} &       & \multicolumn{1}{r}{} \\
    \midrule
    \multicolumn{22}{|c|}{\textbf{JointAutoregressive}} \\
    \midrule
    \multirow{2}[2]{*}{\textbf{Module}} & \multirow{2}[2]{*}{\textbf{Layer}} & \multicolumn{2}{c|}{\textcolor[rgb]{ 1,  0,  0}{\textbf{channel (low)}}} & \multicolumn{2}{c|}{\textcolor[rgb]{ 1,  0,  0}{\textbf{channel (high)}}} & \multicolumn{2}{c|}{\textcolor[rgb]{ 1,  0,  0}{\textbf{ (Q1)}}} & \multicolumn{2}{c|}{\textcolor[rgb]{ 1,  0,  0}{\textbf{(Q2)}}} & \multicolumn{2}{c|}{\textcolor[rgb]{ 1,  0,  0}{\textbf{(Q3)}}} & \multicolumn{2}{c|}{\textcolor[rgb]{ 1,  0,  0}{\textbf{ (Q4)}}} & \multicolumn{2}{c|}{\textcolor[rgb]{ 1,  0,  0}{\textbf{(Q5)}}} & \multicolumn{2}{c|}{\textcolor[rgb]{ 1,  0,  0}{\textbf{ (Q6)}}} & \multicolumn{2}{c|}{\textcolor[rgb]{ 1,  0,  0}{\textbf{(Q7)}}} & \multicolumn{2}{c|}{\textcolor[rgb]{ 1,  0,  0}{\textbf{(Q8)}}} \\
          &       & in    & out   & in    & out   & in    & out   & in    & out   & in    & out   & in    & out   & in    & out   & in    & out   & in    & out   & in    & out  \\
    \midrule
    \multirow{7}[2]{*}{\textbf{ga}} & conv  & 3     & 192   & 3     & 192   & 3     & 30    & 3     & 46    & 3     & 61    & 3     & 80    & 3     & 81    & 3     & 117   & 3     & 130   & 3     & 149 \\
          & GDN   & 192   & 192   & 192   & 192   & 30    & 30    & 46    & 46    & 61    & 61    & 80    & 80    & 81    & 81    & 117   & 117   & 130   & 130   & 149   & 149 \\
          & conv  & 192   & 192   & 192   & 192   & 30    & 66    & 46    & 73    & 61    & 92    & 80    & 122   & 81    & 140   & 117   & 155   & 130   & 174   & 149   & 180 \\
          & GDN   & 192   & 192   & 192   & 192   & 66    & 66    & 73    & 73    & 92    & 92    & 122   & 122   & 140   & 140   & 155   & 155   & 174   & 174   & 180   & 180 \\
          & conv  & 192   & 192   & 192   & 192   & 66    & 56    & 73    & 73    & 92    & 116   & 122   & 99    & 140   & 144   & 155   & 165   & 174   & 178   & 180   & 190 \\
          & GDN   & 192   & 192   & 192   & 192   & 56    & 56    & 73    & 73    & 116   & 116   & 99    & 99    & 144   & 144   & 165   & 165   & 178   & 178   & 190   & 190 \\
          & conv  & 192   & 192   & 192   & 320   & 56    & 192   & 73    & 192   & 116   & 192   & 99    & 192   & 144   & 320   & 165   & 320   & 178   & 320   & 190   & 320 \\
    \midrule
    \multirow{7}[2]{*}{\textbf{gs}} & conv  & 192   & 192   & 320   & 192   & 192   & 101   & 192   & 126   & 192   & 141   & 192   & 163   & 320   & 168   & 320   & 182   & 320   & 176   & 320   & 175 \\
          & GDN   & 192   & 192   & 192   & 192   & 101   & 101   & 126   & 126   & 141   & 141   & 163   & 163   & 168   & 168   & 182   & 182   & 176   & 176   & 175   & 175 \\
          & conv  & 192   & 192   & 192   & 192   & 101   & 49    & 126   & 66    & 141   & 78    & 163   & 93    & 168   & 94    & 182   & 107   & 176   & 127   & 175   & 143 \\
          & GDN   & 192   & 192   & 192   & 192   & 49    & 49    & 66    & 66    & 78    & 78    & 93    & 93    & 94    & 94    & 107   & 107   & 127   & 127   & 143   & 143 \\
          & conv  & 192   & 192   & 192   & 192   & 49    & 59    & 66    & 71    & 78    & 82    & 93    & 105   & 94    & 112   & 107   & 121   & 127   & 107   & 143   & 126 \\
          & GDN   & 192   & 192   & 192   & 192   & 59    & 59    & 71    & 71    & 82    & 82    & 105   & 105   & 112   & 112   & 121   & 121   & 107   & 107   & 126   & 126 \\
          & conv  & 192   & 3     & 192   & 3     & 59    & 3     & 71    & 3     & 82    & 3     & 105   & 3     & 112   & 3     & 121   & 3     & 107   & 3     & 126   & 3 \\
    \midrule
    \multicolumn{1}{r}{} & \multicolumn{1}{r}{} &       & \multicolumn{1}{r}{} &       & \multicolumn{1}{r}{} &       & \multicolumn{1}{r}{} &       & \multicolumn{1}{r}{} &       & \multicolumn{1}{r}{} &       & \multicolumn{1}{r}{} &       & \multicolumn{1}{r}{} &       & \multicolumn{1}{r}{} &       & \multicolumn{1}{r}{} &       & \multicolumn{1}{r}{} \\
    \midrule
    \multicolumn{22}{|c|}{\textbf{Factorizedprior}} \\
    \midrule
    \multirow{2}[2]{*}{\textbf{Module}} & \multirow{2}[2]{*}{\textbf{Layer}} & \multicolumn{2}{c|}{\textcolor[rgb]{ 1,  0,  0}{\textbf{channel (low)}}} & \multicolumn{2}{c|}{\textcolor[rgb]{ 1,  0,  0}{\textbf{channel (high)}}} & \multicolumn{2}{c|}{\textcolor[rgb]{ 1,  0,  0}{\textbf{ (Q1)}}} & \multicolumn{2}{c|}{\textcolor[rgb]{ 1,  0,  0}{\textbf{(Q2)}}} & \multicolumn{2}{c|}{\textcolor[rgb]{ 1,  0,  0}{\textbf{(Q3)}}} & \multicolumn{2}{c|}{\textcolor[rgb]{ 1,  0,  0}{\textbf{ (Q4)}}} & \multicolumn{2}{c|}{\textcolor[rgb]{ 1,  0,  0}{\textbf{(Q5)}}} & \multicolumn{2}{c|}{\textcolor[rgb]{ 1,  0,  0}{\textbf{ (Q6)}}} & \multicolumn{2}{c|}{\textcolor[rgb]{ 1,  0,  0}{\textbf{(Q7)}}} & \multicolumn{2}{c|}{\textcolor[rgb]{ 1,  0,  0}{\textbf{(Q8)}}} \\
          &       & in    & out   & in    & out   & in    & out   & in    & out   & in    & out   & in    & out   & in    & out   & in    & out   & in    & out   & in    & out  \\
    \midrule
    \multirow{7}[2]{*}{\textbf{ga}} & conv  & 3     & 128   & 3     & 192   & 3     & 35    & 3     & 44    & 3     & 58    & 3     & 75    & 3     & 93    & 3     & 93    & 3     & 115   & 3     & 133 \\
          & GDN   & 128   & 128   & 192   & 192   & 35    & 35    & 44    & 44    & 58    & 58    & 75    & 75    & 93    & 93    & 93    & 93    & 115   & 115   & 133   & 133 \\
          & conv  & 128   & 128   & 192   & 192   & 35    & 40    & 44    & 56    & 58    & 75    & 75    & 93    & 93    & 108   & 93    & 151   & 115   & 173   & 133   & 177 \\
          & GDN   & 128   & 128   & 192   & 192   & 40    & 40    & 56    & 56    & 75    & 75    & 93    & 93    & 108   & 108   & 151   & 151   & 173   & 173   & 177   & 177 \\
          & conv  & 128   & 128   & 192   & 192   & 40    & 33    & 56    & 44    & 75    & 55    & 93    & 67    & 108   & 79    & 151   & 134   & 173   & 156   & 177   & 172 \\
          & GDN   & 128   & 128   & 192   & 192   & 33    & 33    & 44    & 44    & 55    & 55    & 67    & 67    & 79    & 79    & 134   & 134   & 156   & 156   & 172   & 172 \\
          & conv  & 128   & 192   & 192   & 320   & 33    & 128   & 44    & 128   & 55    & 128   & 67    & 128   & 79    & 128   & 134   & 320   & 156   & 320   & 172   & 320 \\
    \midrule
    \multirow{7}[2]{*}{\textbf{gs}} & conv  & 192   & 128   & 320   & 192   & 128   & 65    & 128   & 85    & 128   & 102   & 128   & 117   & 128   & 127   & 320   & 177   & 320   & 178   & 320   & 179 \\
          & GDN   & 128   & 128   & 192   & 192   & 65    & 65    & 85    & 85    & 102   & 102   & 117   & 117   & 127   & 127   & 177   & 177   & 178   & 178   & 179   & 179 \\
          & conv  & 128   & 128   & 192   & 192   & 65    & 53    & 85    & 66    & 102   & 72    & 117   & 88    & 127   & 101   & 177   & 120   & 178   & 131   & 179   & 133 \\
          & GDN   & 128   & 128   & 192   & 192   & 53    & 53    & 66    & 66    & 72    & 72    & 88    & 88    & 101   & 101   & 120   & 120   & 131   & 131   & 133   & 133 \\
          & conv  & 128   & 128   & 192   & 192   & 53    & 34    & 66    & 49    & 72    & 67    & 88    & 77    & 101   & 94    & 120   & 124   & 131   & 131   & 133   & 135 \\
          & GDN   & 128   & 128   & 192   & 192   & 34    & 34    & 49    & 49    & 67    & 67    & 77    & 77    & 94    & 94    & 124   & 124   & 131   & 131   & 135   & 135 \\
          & conv  & 128   & 3     & 192   & 3     & 34    & 3     & 49    & 3     & 67    & 3     & 77    & 3     & 94    & 3     & 124   & 3     & 131   & 3     & 135   & 3 \\
    \bottomrule
    \end{tabular}%
  \label{tab:addlabel}%
\end{table}%

\end{document}